\documentclass[mathleft
]{an}
\usepackage{aas_macros} 
\usepackage{graphicx}
\usepackage{times}
\usepackage{natbib}
\bibpunct{(}{)}{;}{a}{}{,}
\begin{document}

\Pagespan{789}{}
\Yearpublication{2006}%
\Yearsubmission{2005}%
\Month{11}%
\Volume{999}%
\Issue{88}%

\title{Solar-like oscillations in cluster stars\thanks{Data
from \it{Kepler}}}

\author{D. Stello\inst{1}\fnmsep\thanks{Corresponding author:
  \email{stello@physics.usyd.edu.au}\newline}
\and S.~Basu\inst{2}
\and T.~R.~Bedding\inst{1}
\and K.~Brogaard\inst{3}
\and H.~Bruntt\inst{4}
\and W.~J.~Chaplin\inst{5} 
\and J.~Christensen-Dalsgaard\inst{3} 
\and P.~Demarque\inst{2} 
\and Y.~P.~Elsworth\inst{5} 
\and R.~A.~Garc{\'\i}a\inst{6} 
\and R.~L.~Gilliland\inst{7} 
\and S.~Hekker\inst{5} 
\and D.~Huber\inst{1}
\and C.~Karoff\inst{5} 
\and H.~Kjeldsen\inst{3} 
\and Y.~Lebreton\inst{8}
\and S.~Mathur\inst{9}
\and S.~Meibom\inst{10} 
\and J.~Molenda-\.Zakowicz\inst{11} 
\and A.~Noels\inst{12} 
\and I.~W.~Roxburgh\inst{13} 
\and V.~S.~Aguirre \inst{14}
\and C.~Sterken\inst{15} 
\and R.~Szab\'o\inst{16} 
}
\titlerunning{Solar-like oscillations in cluster stars}
\authorrunning{D. Stello et al.}
\institute{
Sydney Institute for Astronomy (SIfA), School of Physics, University of Sydney, NSW 2006, Australia
\and 
Department of Astronomy, Yale University, P.O. Box 208101, New Haven, CT 06520-8101
\and 
Department of Physics and Astronomy, Aarhus University, 8000 Aarhus C, Denmark
\and 
LESIA, CNRS, Universit\'e Pierre et Marie Curie, Universit\'e Denis Diderot, Observatoire de Paris, 92195 Meudon, France
\and 
School of Physics and Astronomy, University of Birmingham, Edgbaston, Birmingham B15 2TT, UK
\and 
Laboratoire AIM, CEA/DSM-CNRS, Universit\'e Paris 7 Diderot, IRFU/SAp, Centre de Saclay, 91191, Gif-sur-Yvette, France
\and 
Space Telescope Science Institute, 3700 San Martin Drive, Baltimore, Maryland 21218, USA
\and 
GEPI, Observatoire de Paris, CNRS, Universit\'e  Paris Diderot, 5 Place Jules Janssen, 92195 Meudon, France
\and 
High Altitude Observatory, NCAR, P.O. Box 3000, Boulder, CO 80307, USA
\and 
Harvard-Smithsonian Center for Astrophysics, 60 Garden Street, Cambridge, MA, 02138, USA
\and 
Instytut Astronomiczny Uniwersytetu Wroc\l{}awskiego, ul.\ Kopernika 11, 51-622 Wroc\l{}aw, Poland
\and 
Institut d'Astrophysique et de G\'eophysique de l'Universit\'e de Li\`ege, 17 All\'ee du 6 Ao\^ut, B-4000 Li\`ege, Belgium
\and 
Queen Mary University of London, Mile End Road, London E1 4NS, UK
\and 
Max Planck Institute for Astrophysics, Karl Schwarzschild Str. 1, Garching bei M\"{u}nchen, D-85741, Germany
\and 
Vrije Universiteit Brussel, Pleinlaan 2, B-1050 Brussels, Belgium
\and 
Konkoly Observatory, H-1525 Budapest, P.O. Box 67, Hungary
}

\received{25 May 2010}
\accepted{24 June 2010}
\publonline{later}


\abstract{
  We present a brief overview of the history of attempts to
  obtain a clear detection of solar-like oscillations in cluster stars, and
  discuss the results on the first clear detection, which was made by the
  Kepler Asteroseismic Science Consortium (KASC) Working Group 2. 
}

\maketitle

\section{Introduction}
Star clusters are  extremely important in stellar astrophysics.  Most stars
form in open clusters, many of which disperse into the diversity of field
stars in the interstellar medium.  Understanding the formation and 
evolution of cluster stars is therefore important for achieving a
comprehensive theory of stellar evolution.  Stars in a cluster
are thought to be formed coevally, from the same interstellar cloud of gas
and dust.  Each cluster member is therefore expected to have some properties
in common (age, composition, distance), which strengthens our ability to
constrain our stellar models when tested against an ensemble of cluster
stars, especially for asteroseismic analyses \citep{GoughNovotny93}. 
Asteroseismology has the capability to probe the interior of stars
and hence help us understand the fundamental physical process that govern
stellar structure and evolution \citep[e.g.,][]{Dalsgaard02}.  
In particular, the detection of solar-like oscillations provide many modes,
which each carrying unique information about the stellar interior.
Stars that potentially exhibit solar-like oscillations,
covering most stars that we see, are
cooler than the red edge of the classical instability strip, and have a
convection zone near the surface (necessary for the excitation of the modes).
Solar-like oscillations are reasonably well described by current theory,
giving us some
confidence that we can use them as tools to understand stellar physics, and
hopefully also to learn more about the more subtle aspects of the
oscillations themselves. 
Combining asteroseismic analysis of solar-like oscillations with the study
of cluster stars has therefore been a long-sought goal. 

\section{Previous attempts}
\begin{figure}
\includegraphics[width=80mm]{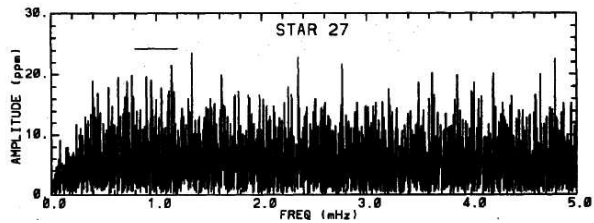}
\caption{Amplitude spectrum (high-pass filtered) of one of the stars
  targeted by \citet{Gilliland93}. The horizontal line marks the expected
  location of the oscillations. 
        }
\label{fig1}
\end{figure}

\begin{figure}
\includegraphics[width=80mm]{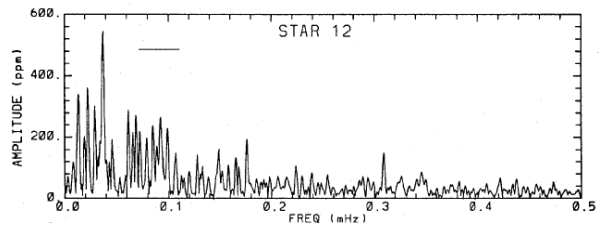}
\caption{Amplitude spectrum of red giant star observed by
  \citet{Gilliland93}. Horizontal line marks the expected 
  location of the oscillations. 
        }
\label{fig2}
\end{figure}

{\it Kepler} is certainly not the first attempt to detect
solar-like oscillations in cluster stars.
A quick  (and hence incomplete) perusal of the history of
previous attempts to detect solar-like oscillations in open and globular
clusters shows that   
several attempts were made to detect oscillations since 
the early 1990s.  Among the most ambitious was that of
\citet{Gilliland93}, who used 4-m class telescopes to target the stars in
the open cluster M67 at
the cluster turn-off in a multi-site campaign that lasted one week.  While
an impressively low noise level was obtained, the data did not reveal the clear
detection of stellar oscillations (Figure~\ref{fig1}).  However, a red giant
star that happened to be in the field did show intriguing evidence of excess
power in the expected frequency range (Figure~\ref{fig2}).  Unfortunately,
the length of the time series did not allow individual modes to be resolved
for such an evolved star with much smaller frequency separations between 
modes. A clear detection remained elusive, as oscillations could not be
distinguished from the rising background towards low frequency. 


\begin{figure}
\includegraphics[width=80mm]{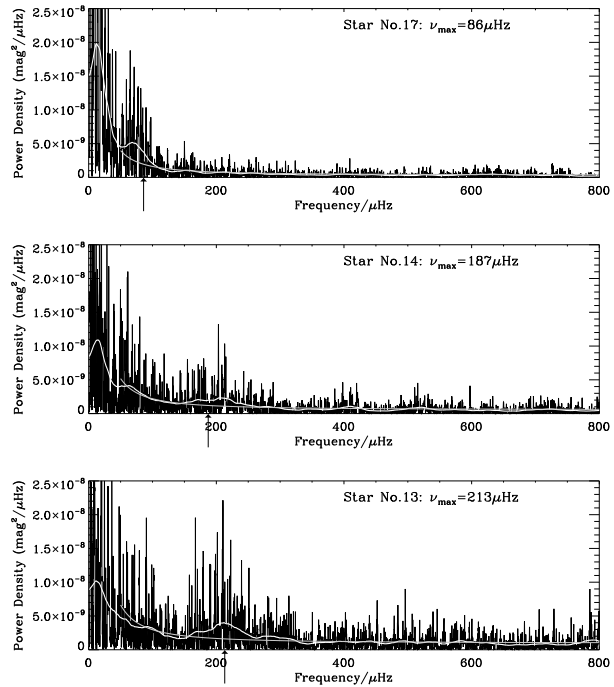}
\caption{Power spectra of three red giant stars observed by
  \citet{Stello07}. Black arrow marks the location of the oscillations
  expected from scaling the solar value. 
        }
\label{fig3}
\end{figure}

Inspired by Gilliand's results, \citet{Stello07} targeted specifically the red
giants in M67 during a 6-week long multi-site campaign of 1--2m class telescopes.  
Strong evidence for excess power was found in a number of stars, but no
unambiguous detection of the solar-like pattern of equally spaced modes was
claimed by the authors (Figure~\ref{fig3}). 

In parallel, several attempts to detect oscillations in globular clusters
were carried out. From the ground, \citet{Frandsen07} aimed at the red giants
in M4, which delivered lower limits on amplitudes, indicating that the
low metallicity of M4 could have the effect of lowering the oscillation
amplitudes. Again, detection was hindered by long-term stability not being
high enough and varying data quality resulting in strong aliasing in the
weighted amplitude spectra. 
Slightly more successful were the efforts using the {\it Hubble Space
Telescope} by \citet{EdmondsGilliland96,StelloGilliland09}. 
In the former study,
clear variation was found in a large number of red giants in 47 Tuc, but the
low frequency resolution provided by the 40-hour time series did
not allow the authors to establish this as solar-like oscillations.  The
later study was aimed at the red giants in the extremely metal poor
NGC 6397, using archival data originally obtained to detect the cluster's
faint white dwarf population. The far from ideal 
data of highly saturated photometry of the red giants meant that only one
star showed good evidence for oscillations, with excess power at the right
frequency range and amplitude.  Despite the 27-day long time
series, this fell just short for an unambiguous detection of equally
spaced frequencies in this highly evolved asymptotic giant branch star.

The main conclusion from these previous efforts is that dedicated
space-based missions are required to achieve the ultra-high precision
photometry and long-term stability in order to detect
solar-like oscillations in clusters with such accuracy that they will be
useful for asteroseismic analysis.

We note that in addition to the previous marginal detections, these
campaigns resulted in firm detection of oscillations in 
a number of classical pulsators that exhibit much large amplitude than
solar-like oscillations \citep[see e.g.][~and references
therein]{Bruntt07}.

\section{First results from \it{Kepler}}
{\it Kepler} has a unique capability to overcome the shortcomings that
have limited previous efforts aimed at stellar clusters.  Both quality and
quantity of the{ \it Kepler} data outshine that of early explorations by
several orders of magnitude, and it will undoubtedly be the front runner for
cluster seismology in the next 5--10 years.

\begin{figure}
\includegraphics[width=80mm]{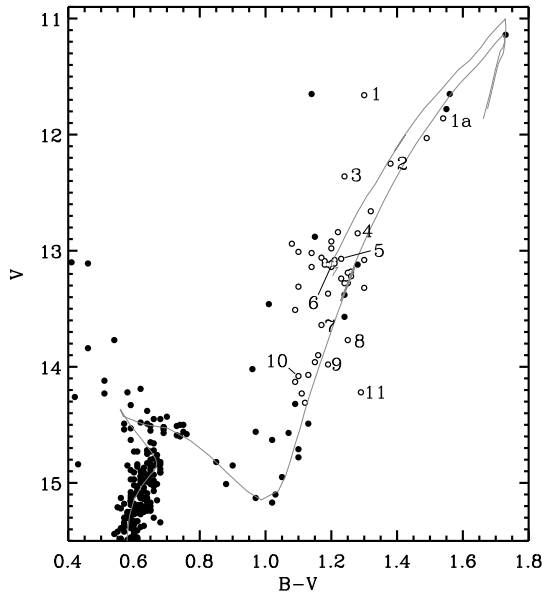}
\caption{HR-diagram of NGC 6819. Empty symbols mark those where a detection
  of solar-like oscillations was reported by \citet{Stello10}.
        }
\label{hrd}
\end{figure}

\begin{figure}
\includegraphics[width=80mm]{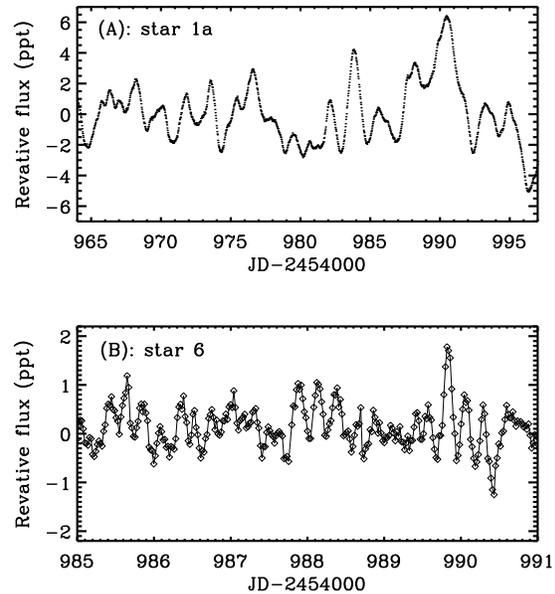}
\caption{{\it Kepler} time series for two red giants in NGC 6819. Numbers
  refer to the numbering in Figure~\ref{hrd}. Note the different time scale
  of the variation. Photometry and isochrone is of \citet{Hole09} and
  \citet{Marigo08}, respectively.
        }
\label{f1a}
\end{figure}

\begin{figure}
\includegraphics[width=80mm]{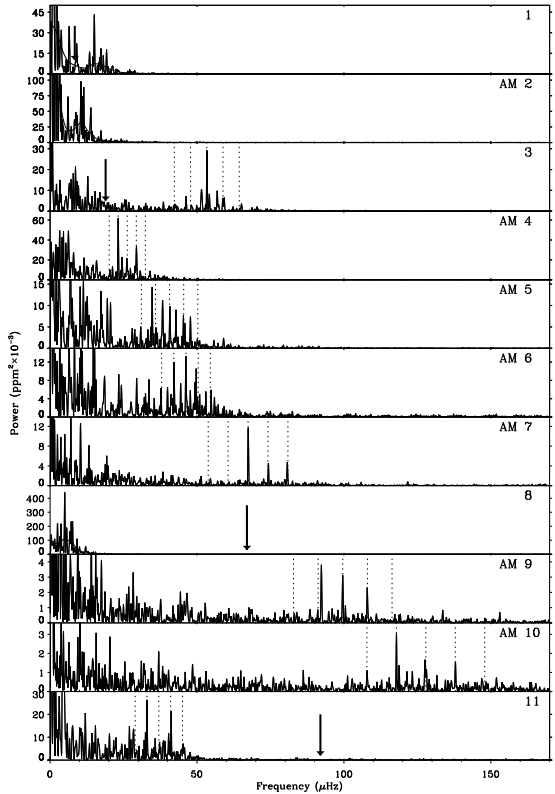}
\caption{Power spectra of 11 stars marked in Figure~\ref{hrd}, which
  are representative for the entire sample. `AM' indicates that the star is
  an asteroseismic member (i.e. observation agrees with scaling
  relations). Dashed lines show the measured large frequency
  separation. For stars where the large separation could not be determined
  (no dashed lines), 
  we localised the power excess from the hump of power in the smoothed
  power spectrum (solid black curve).  The arrows indicate the expected
  location of the excess power for stars where observations do not agree
  with expectation.
        }
\label{spectra}
\end{figure}

As reported by \citet{Stello10}, the first month of {\it Kepler} data
already revealed clear detection of solar-like oscillations in a large
sample of red giant stars in the open cluster NGC 6819 \citep[see also][]{Gilliland10}. 
Based on the spacecrafts so called long-cadence mode, which provides a time 
averaged exposure every 29.4 minutes, detection was reported in 47 red
giant stars that range almost from the bottom to the tip of the red giant
branch (Figure~\ref{hrd}). 
We saw periodicity in the light curves that
span about a factor of 100, corresponding to a factor of $\sim 10$ in
radius. Two sample light curves are shown in Figure~\ref{f1a}.
Power spectra of the stars marked with numbers in
Figure~\ref{hrd} are shown in Figure~\ref{spectra}. Panels are sorted
according to apparent magnitude (brightest at the top), which for a cluster is indicative of
luminosity.  One noticeable result is that not all
stars with high membership probability from radial velocity surveys
\citep[see][]{Hole09} follow the expected monotonic trend of increasing
frequency of the oscillations (and decreasing amplitude) for decreasing
luminosity. We indicate the expected frequency location with an arrow for
stars that seem to behave strangely compared to the classical scaling
relations for the amplitude and the frequency of maximum power
\citep[e.g.][]{KjeldsenBedding95}.   
Possible explanations for this behaviour are that these ``odd'' stars are not
members, or that they have unusual evolution histories.

\citet{Stello10} were further able to measure the amplitudes of the modes using the
method by \citet{Kjeldsen08}, assuming the relative amplitudes of the modes
of different spherical degree was the same as for the Sun. From this we
could test the $L/M$ scaling relation \citep{KjeldsenBedding95,Samadi07},
and found that $(L/M)^{0.7}/T_{\mathrm{eff}}^2$ provided the best match to
the data. 

For further details on what is reported here, we refer to the source paper
of \citet{Stello10}.

\section{Future}
There are four open clusters in {\it Kepler's} field of view. They span a
range in metallicity and age, which brackets the solar values, and are
therefore ideal for testing our current models of stellar evolution
(Table~\ref{tab}). 

\begin{table}
\caption{Open clusters in {\it Kepler} field}
\label{tab}
\begin{tabular}{llll}\hline
Cluster & Age & [Fe/H] & M$_\mathrm{turnoff}$\\ 
        & Gyr &        & M$_\odot$           \\ 
\hline
NGC 6866 & $\sim$0.4 & $\sim -$0.1  & $\sim$1.7 \\
NGC 6811 & $\sim$1.0 & $\sim -$0.07 & $\sim$1.5 \\
NGC 6819 & $\sim$2.5 & $\sim -$0.05 & $\sim$1.3 \\
NGC 6791 & $\sim$8.5 & $\sim +$0.4  & $\sim$1.0 \\
\hline
\multicolumn{4}{l}{\scriptsize{Values are from \citet{Grundahl08} (NGC
        6791),}}\\
\multicolumn{4}{l}{\scriptsize{ \citet{Hole09} (NGC 6819),
        \citet{LoktinMatkin94} (NGC 6866)}}\\
\multicolumn{4}{l}{\scriptsize{ and unpublished work by Meibom.}}
\end{tabular}
\end{table}

\begin{figure}
\includegraphics[width=80mm]{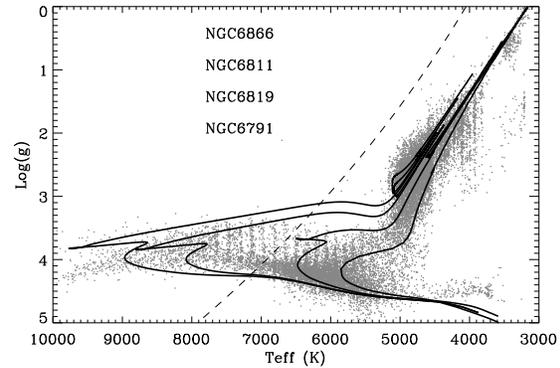}
\caption{$\log(g$) vs $T_{\mathrm{eff}}$ for stars in {\it Kepler's} field
  of view. We represent the four open clusters by suitable
  isochrones. The order in which we have plotted the cluster names
  corresponds to their turn-off stars, with NGC 6866 having the hottest (heaviest)
  turn-off stars and NGC 6791 the coolest (lightest). The dashed line
  indicates the red edge of the classical instability strip.
        }
\label{kic}
\end{figure}

In Figure~\ref{kic} we show $\log(g$) vs $T_{\mathrm{eff}}$ for a
representative sample of the stars in {\it Kepler's} field of view together
with the representative isochrones for the four open clusters that are
targets in our future asteroseismic analyses.

For NGC 6819 we expect to achieve a signal-to-noise level for the turn-off
stars that after 3.5 years of data matches what we see in the bottom panels
of Figure~\ref{spectra}.  This will provide detection in up to 100 stars
ranging stellar evolution from the main sequence F stars to the asymptotic
giant branch including M giants, as well as a number of blue
stragglers. This will potentially provide unprecedented tests of
state-of-the-art stellar evolution models.  

In NGC 6791 we already see evidence for power in the red giants, and
expect firm detections for all stars on this highly populated red giant
branch, with unique potential for testing intrinsic variation among
practically identical stars.

The two younger clusters NGC 6811 and NGC 6866 are less populated but
provide the opportunity to investigate classical pulsators in
great detail. NGC 6811 also contains a few He-core burning red giants.

The combination of results from all four clusters promises great prospects
for testing asteroseismic scaling relations on distinct stellar populations
that span a large range in stellar age and brackets the solar metallicity.

\acknowledgements
Funding of the Discovery mission is provided by NASA's Science Mission
Directorate. The authors thank the entire {\it Kepler} team without whom
this investigation would not have been possible. The authors also thank all
funding councils and agencies that have supported the activities for
Working Group 2 of the KASC. In particular, DS would like to thank HELAS for
support to attend the HELAS IV meeting in Lanzarote.

\bibliography{bib_complete}

\end{document}